\documentclass[twocolumn,preprintnumbers,amsmath,amssymb,nofootinbib]{revtex4}

\usepackage{graphicx}
\usepackage{dcolumn}
\usepackage{bm}
\usepackage{amsthm}

\newcommand{\glog}{\Lambda}
\newcommand{\gexp}{{\cal E}}

\theoremstyle{definition}

\theoremstyle{remark}

\begin{document}


\title{A comprehensive classification of complex statistical systems and an ab-initio derivation of their entropy and distribution functions}

\author{Rudolf Hanel$^{1}$}
\email{rudolf.hanel@meduniwien.ac.at}
\author{Stefan Thurner $^{1,2}$}
\email{stefan.thurner@meduniwien.ac.at}

\affiliation{ 
   $^{1}$ Section for Science of Complex Systems; Medical University of Vienna; Spitalgasse 23; 1090 Vienna; Austria \\
   $^{2}$ Santa Fe Institute; 1399 Hyde Park Road; Santa Fe; NM 87501; USA\\
      }
 
\begin{abstract}
To characterize strongly interacting statistical systems within a thermodynamical framework -- complex systems in particular -- 
it might be necessary to introduce generalized entropies, $S_g$. 
A series of such entropies have been proposed in the past, mainly to accommodate important empirical 
distribution functions to a maximum ignorance principle. 
Until now the understanding of the fundamental origin of these entropies and its 
deeper relations to complex systems is limited.  
Here we explore this questions from first principles.  We start by observing that the 4th Khinchin axiom (separability axiom)
is violated by strongly interacting systems in general and 
ask about the consequences of violating the 4th axiom while assuming the first three  Khinchin axioms (K1-K3) to hold  and $S_g=\sum_ig(p_i)$. 
We prove by simple scaling arguments that under these requirements {\em each} statistical system 
is uniquely characterized by a distinct pair of scaling exponents $(c,d)$ in the large size limit. 
The exponents  define equivalence classes for  all interacting and non interacting systems. 
This allows to derive a unique entropy, 
$S_{c,d}\propto \sum_i \Gamma(d+1, 1- c \ln p_i)$,
which covers all entropies which respect K1-K3 and can be written as $S_g=\sum_ig(p_i)$.
Known entropies  can now be  classified within these equivalence classes.
The corresponding distribution functions are special forms of Lambert-$W$ exponentials containing  as  special cases  
Boltzmann,  stretched exponential and Tsallis distributions (power-laws) -- all widely abundant in nature. 
This is, to our knowledge, the first {\em ab initio} justification for the existence of generalized entropies.  
Even though here we assume $S_g=\sum_ig(p_i)$, we show that more general entropic forms can be classified along the same lines. 
\end{abstract}

\pacs{
}  


\maketitle

Weakly interacting statistical systems can be perfectly described by thermodynamics -- provided the number of states  $W$ in the system is large.
Complex systems in contrast, characterized by long-range  and strong interactions, can fundamentally change their macroscopic
qualitative properties as a function of the number of states or  the  degrees of freedom. 
This leads to the extremely rich behavior of complex systems when compared to simple ones, such as gases.  
The need for understanding the macroscopic properties of such interacting systems on the basis of a few
measurable quantities only, 
is reflected in      
the hope that a thermodynamic approach can also be 
established for interacting systems.
In particular it is hoped that appropriate entropic forms can be found for specific systems 
at hand, which under the assumption of maximum ignorance, could explain 
sufficiently stationary macro states of these systems.  
In this context a series of entropies have been suggested over the past decades, 
\cite{tsallis88, celia, kaniadakis,curado,expo_ent,ggent} and Table 1. 
So far the origin of such entropies has not been fully understood within a general framework .   

Here we propose a  general classification scheme of both interacting (complex) 
and non- or weakly-interacting statistical systems in 
terms of their asymptotic behavior under changes of the number of degrees of freedom of the system.
Inspired by the classical works of Shannon \cite{shannon} and Khinchin \cite{kinchin_1} 
we follow a classical scaling approach 
to study systems where 
the first three Khinchin axioms hold;  we study the 
consequences of the violation of the fourth, which is usually referred to as the separation axiom. The first 
3 Khinchin axioms are most reasonable to hold also in strongly interacting systems. 

The central concept in understanding macroscopic system behavior on the basis of 
microscopic properties is {\em entropy}. 
Entropy relates the number of states of a system to an {\em extensive} quantity, which plays a fundamental role in the systems thermodynamical description. Extensive means that if two initially isolated, 
i.e. sufficiently separated 
systems, $A$ and $B$, with $W_A$ and $W_B$ the respective numbers of states, are brought together, 
the entropy of the combined system 
$A+B$ is $S(W_{A+B}) = S(W_A) + S(W_B)$. 
$W_{A+B}$ is the number of states in the combined system $A+B$.
This is not to be confused with {\em additivity} which is the property that $S(W_A W_B) = S(W_A) + S(W_B)$. 
Both, extensivity and additivity coincide if  number of states in the combined system is $W_{A+B}=W_AW_B$.
Clearly,  for a non-interacting system Boltzmann-Gibbs entropy, $S_{\rm BG}[p]=\sum_i g_{\rm BG}(p_i)$,  
with $g_{\rm BG}(x)=- x\ln x$, is extensive and additive.
By 'non-interacting'  (short-range, ergodic, sufficiently mixing, Markovian, ...) systems we mean  $W_{A+B}=W_AW_B$. 
For interacting statistical systems the latter is in general not true; phase space is only partly visited and $W_{A+B} < W_AW_B$. In this case,   
an additive entropy  such as Boltzmann-Gibbs can no longer be  extensive and vice versa. 
To keep the possibility to treat interacting statistical systems with a thermodynamical formalism 
and to ensure extensivity of entropy, a proper  entropic form must be found for 
the particular interacting statistical systems at hand. 
We call these entropic forms  {\em generalized entropies}  and assume them to be of the form 
\begin{equation}
 S_g[p]=\sum_{i=1}^W g(p_i) \quad ,
\label{S_g} 
\end{equation} 
$W$ being the number of states\footnote{\label{fooSum} 
   Obviously not all generalized entropic forms are of this type. R\'enyi entropy e.g. is  
   of the form $G(\sum_{i} g(p_i))$, with $G$ a monotonic function.
   We use the entropic forms Eq. (\ref{S_g}) for simplicity and for their nice characterization in terms of asymptotic properties.
   Using the R\'enyi form, many asymptotic properties can be studied in exactly the same way as will be shown here, however it gets
   technically more involved since asymptotic properties of $G$ and $g$ have to be dealt with simultaneously.
}.
The four Khinchin axioms (K1-K4) uniquely determine $g$ to be the Boltzmann-Gibbs-Shannon (BG) entropy \cite{kinchin_1}.
These axioms have a series of implications on $g$:
\begin{itemize}
\item
K1: The requirement that $S$ depends continuously on $p$  implies that $g$ is a continuous function.
\item
K2: The requirement that the entropy is maximal for the equi-distribution $p_i=1/W$ implies that $g$ is a concave function (for the exact formulation needed in a proof below, see SI Proposition 1).
\item 
K3: The requirement that adding a zero-probability state to a system, $W+1$ with $p_{W+1}=0$,  does not change the entropy, implies $g(0)=0$. 
\item 
K4: The entropy of a system -- split into sub-systems $A$ and $B$ -- equals the entropy of $A$ plus the expectation value of the entropy 
of $B$, conditional on $A$. 
\end{itemize}
If K1 to K4 hold, the entropy is the Boltzmann-Gibbs-Shannon entropy, 
\begin{equation}\label{S_BG} 
S_{\rm BG}[p]=\sum_{i=1}^W g_{\rm BG}(p_i)\quad {\rm with} \quad g_{\rm BG}(x)=-x\ln x .
\end{equation} 
The separability requirement of K4 corresponds exactly to Markovian processes and
is obviously violated for most interacting systems. For these systems, introducing 
generalized entropic forms $S_g[p]$, is one possibility to ensure extensivity of entropy.
We assume in this paper that axioms K1, K2, K3  hold, i.e. we restrict ourselves to $S_g=\sum_i g(p_i)$ with $g$ continuous, concave 
and $g(0)=0$.
These systems we call {\em admissible} systems.

In the following 
we classify all (large) statistical systems where K1-K3 hold   
in terms of two asymptotic properties of their associated generalized entropies. 
Both properties are associated with one scaling function each. 
Each scaling function is characterized by one exponent, $c$ for the first and $d$ for the second property. 
They exponents  allow to define equivalence relations of entropic forms, i.e. two entropic forms are 
equivalent iff their exponents are the same. The pair $(c,d)$ uniquely defines an equivalence class of entropies. 
Each admissible system approaches one of these equivalence classes in its $W\to\infty$ limit. 

To be very clear, by asymptotic we mean the number of states being large, $W \gg 1$.  
Thus all the relevant entropic information on the system is encoded in the properties of $g(x)$ near zero, i.e. in the region 
$x\sim W^{-1}$. 
In the asymptotic limit it is therefore not necessary to know  $g$ on the entire interval of  $x\in [0,1]$, but it is sufficient  to know it in the vicinity of $x\sim 0$. 
In other words  the part of $g(x)$ where $x>W^{-1}$  contains information which is irrelevant for  the macroscopic properties.
In terms of distribution functions this  simply means that everything but the tails becomes irrelevant for large systems.  
This implies that the equivalence classes $(c,d)$ can be interpreted as {\em basins of attraction}
for systems that may differ on small scales but start to behave identical 
in the thermodynamic limit.

We show that a single two-parameter family of entropies  
$S_{c,d}\propto\sum_i  \Gamma(d+1, 1- c \ln p_i)$, 
is sufficient to cover all admissible systems; 
i.e. all entropies of the form $S_g=\sum_i g(p_i)$ are equivalent to some representative
entropy $S_{c,d}$, which parametrizes the equivalence classes $(c,d)$. 
Distribution functions associated with  $S_{c,d}$ involve Lambert-$W$ exponentials. 
Lambert-$W$ functions have deep connections to 
self-similarity and time-delayed differential equations, see e.g. \cite{corless,banwell}.
Important special cases of these distributions are power-laws (Tsallis entropy \cite{TsallisBook_2009}) 
and stretched exponential distributions which are widely abundant in nature.  

\begin{figure}[t]
 \begin{center}
 	\includegraphics[width=\columnwidth] {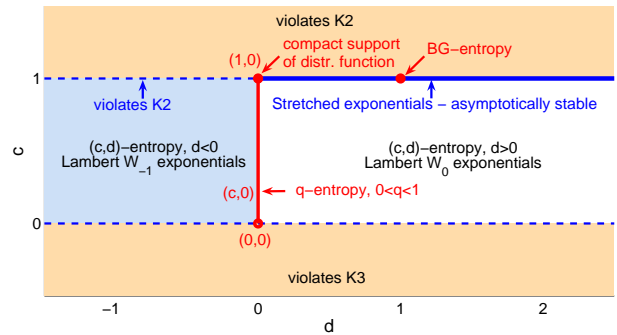}
 \end{center}
\caption{
Equivalence classes of functions $g_{c,d}$ parametrized in the $(c,d)$-plane, with their associated entropies and characteristic distribution functions. 
Following \cite{ggent,HT_hilhorst} entropies are one-to-one related to distribution functions.  
BG entropy corresponds to $(1,1)$, Tsallis entropy to $(c,0)$, and entropies for stretched exponentials to $(1,d>0)$.
All entropies leading to distribution functions with compact support, 
belong to equivalence class $(1,0)$. An example are $S_q$ entropies with $q>1$ (using the maximum entropy principle with usual expectation values in the constraints \cite{ggent,HT_hilhorst}).
\label{classfig}
}
\end{figure}

\section{Asymptotic properties of  non-additive entropies}

We now discuss 2 scaling properties of generalized entropies of the form $S=\sum_i g(p_i)$ assuming the validity of  the first 3 Khinchin axioms. 

The first asymptotic property is found from the scaling relation 
\begin{equation}
 \frac{S_g(\lambda W)}{S_g(W)}=\lambda \frac{g( \frac{1}{\lambda W} )}{g(\frac 1W)}\quad , 
\label{Stog}
\end{equation}
in the limit $W\to\infty$, i.e. by defining the scaling function 
\begin{equation}
   f(z)\equiv\lim_{x\to 0}\frac{g(z x)}{g(x)}    \quad \quad  (0<z<1) \quad. 
\label{f_funct}
\end{equation}
The scaling function $f$ for systems satisfying K1,K2, K3, but not K4, 
 can only be a power $f(z)=z^c$, with $0<c\leq 1$,  given $f$ being continuous. 
This is shown in the SI (Theorem 1). 
Inserting Eq. (\ref{f_funct}) in Eq. (\ref{Stog}) gives the first asymptotic law 
\begin{equation}
 \lim_{W \to \infty }\frac{S_g(\lambda W)}{S_g(W)}=  \lambda ^{1-c} \quad . 
\label{Stog2}
\end{equation}
From this it is clear that 
\begin{equation}
    \lim_{W\to \infty}   \frac{S(\lambda W)}{S(W)}  \lambda^{c-1} =1 \quad.  
\label{f_funct2}
\end{equation}
If we substitute  $\lambda$ in Eq. (\ref{f_funct2}) by $\lambda \to W^a$
we can identify  a second asymptotic property. We define $h_c(a)$
\begin{equation}
  h_c(a)\equiv    \lim_{W\to \infty}   \frac{S(W^{1+a})}{S(W)}  W^{a(c-1)}  = \lim_{x\to 0}  \frac{g(x^{1+a})}{ x^{ac}g(x)} \quad, 
\label{fr_funct}
\end{equation}
with $x=1/W$. 
$h_c(a)$ in principle depends  on $c$ and $a$. 
It can be  proved  (SI, Theorem 2) that $h_c(a)$ is given by  
\begin{equation}
  h_c(a)=(1+a)^d \qquad    (d \,\,\, {\rm constant}) \quad. 
\end{equation}
Remarkably, $h_c$ does not explicitly depend on $c$ anymore and $h_c(a)$ is an asymptotic property which is 
{\em independent}  of the one given in Eq. (\ref{Stog2}). 
Note that if $c=1$, concavity of $g$ implies $d\geq 0$.   \\

\section{Classification of statistical systems}

We are now in the remarkable position to characterize {\em all} large K1-K3 systems by a pair of two exponents $(c,d)$, i.e. 
their scaling functions $f$ and $h_c$. See Fig. \ref{classfig}.

For example, for $g_{\rm BG}(x)=-x\ln(x)$ we have $f(z)=z$, i.e.  $c=1$, 
and $h_{c}(a)=1+a$, i.e.  $d=1$.
$S_{\rm BG}$ therefore belongs to the universality class $(c,d)=(1,1)$.  
For $g_{q}(x)= (x-x^q)/(1-q)$ (Tsallis entropy) and $0<q<1$ one finds 
$f(z)=z^q$, i.e.  $c=q$ and $h_{c}(a)=1$, i.e.  $d=0$, 
and Tsallis entropy, $S_{q}$, belongs to the universality class $(c,d)=(q,0)$. 
A series of other examples are listed in Table 1. 

The universality classes $(c,d)$ are equivalence classes with the 
equivalence relation given by:  
$g_{\alpha} \equiv g_{\beta} \Leftrightarrow c_\alpha=c_\beta$ and $d_\alpha=d_\beta$.
This equivalence relation partitions the space of all admissible $g$ into equivalence classes completely specified by the pair $(c,d)$.  

\begin{figure}[t]
 \begin{center}
	\includegraphics[width=0.8\columnwidth] {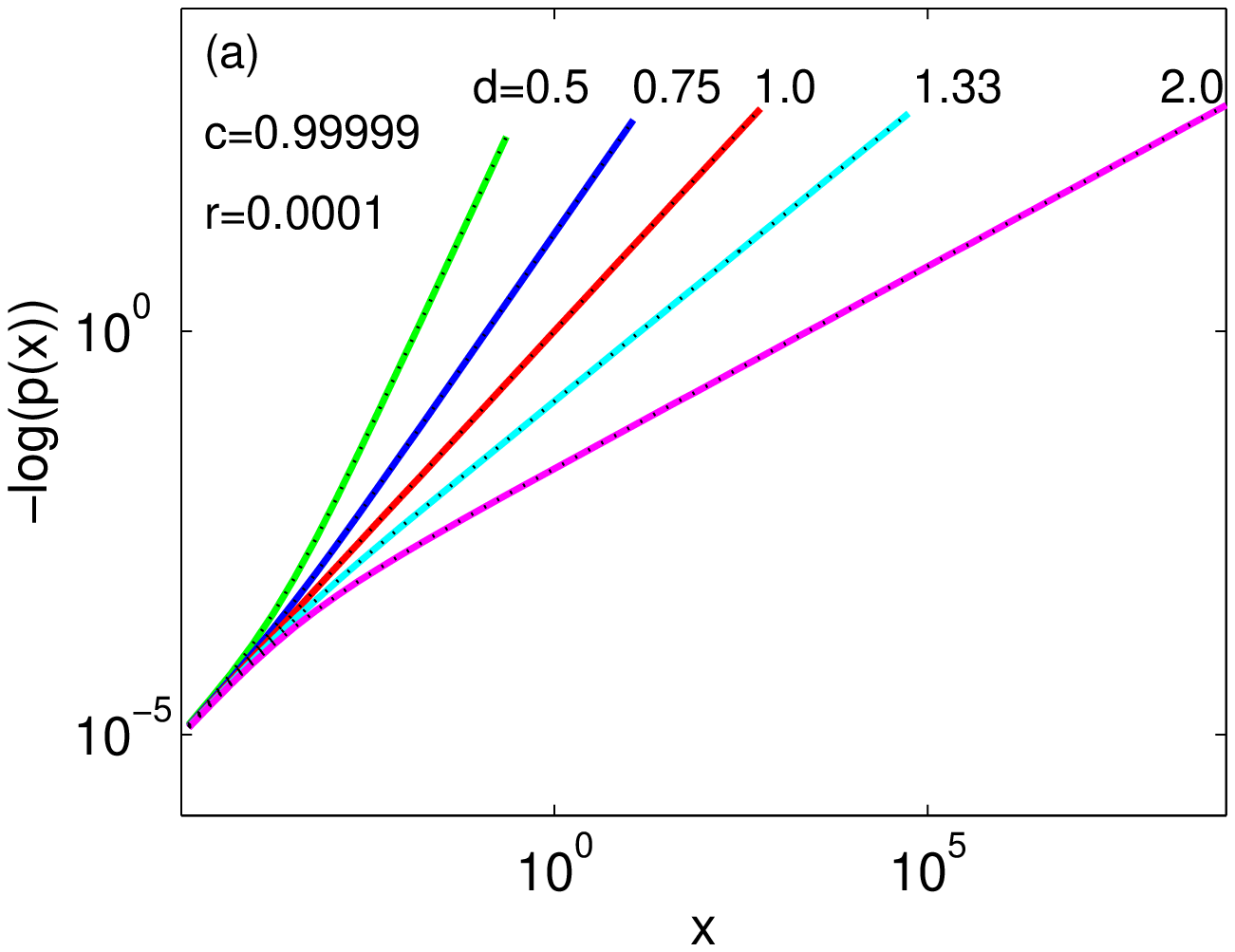}
	\includegraphics[width=0.8\columnwidth] {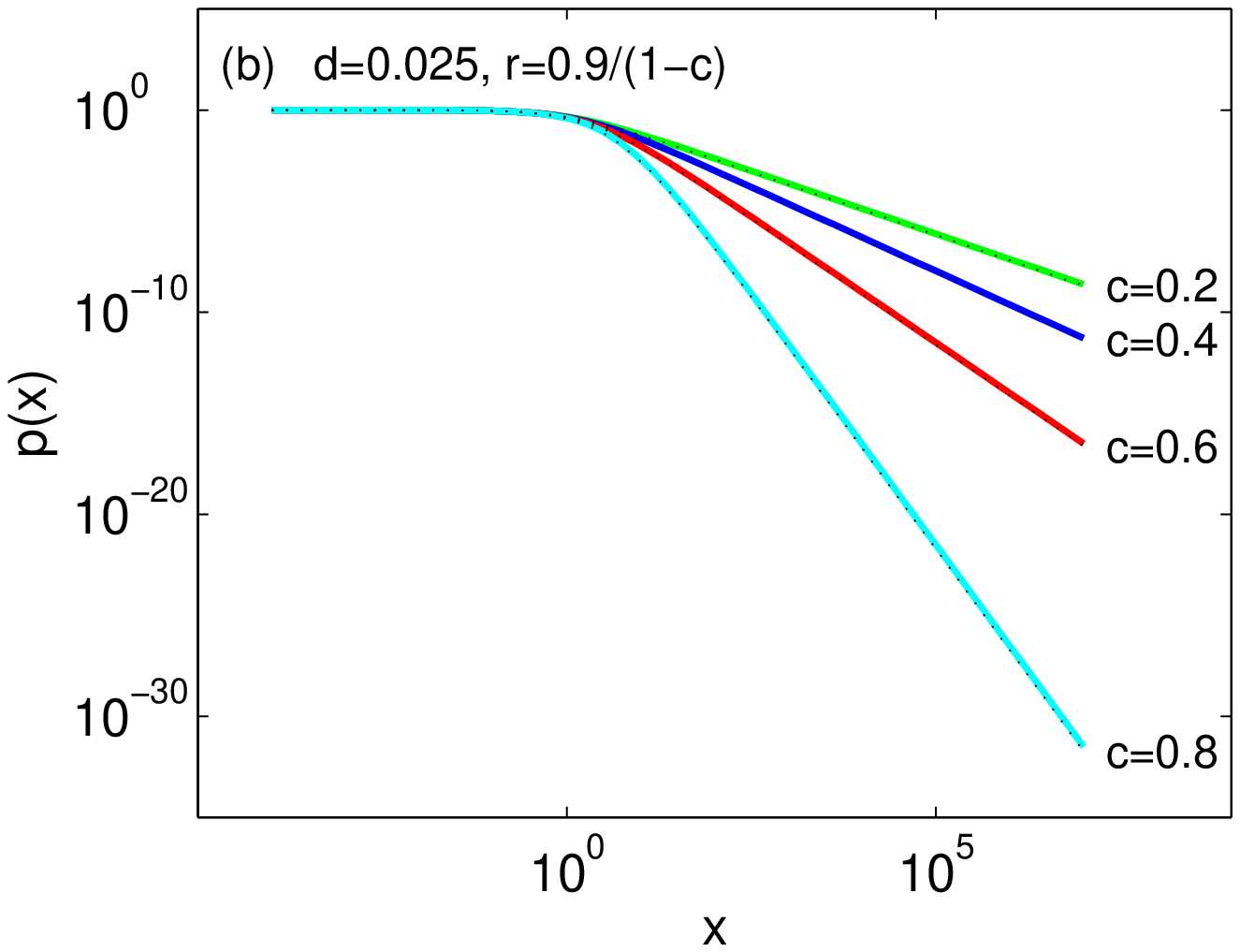}
	\includegraphics[width=0.8\columnwidth] {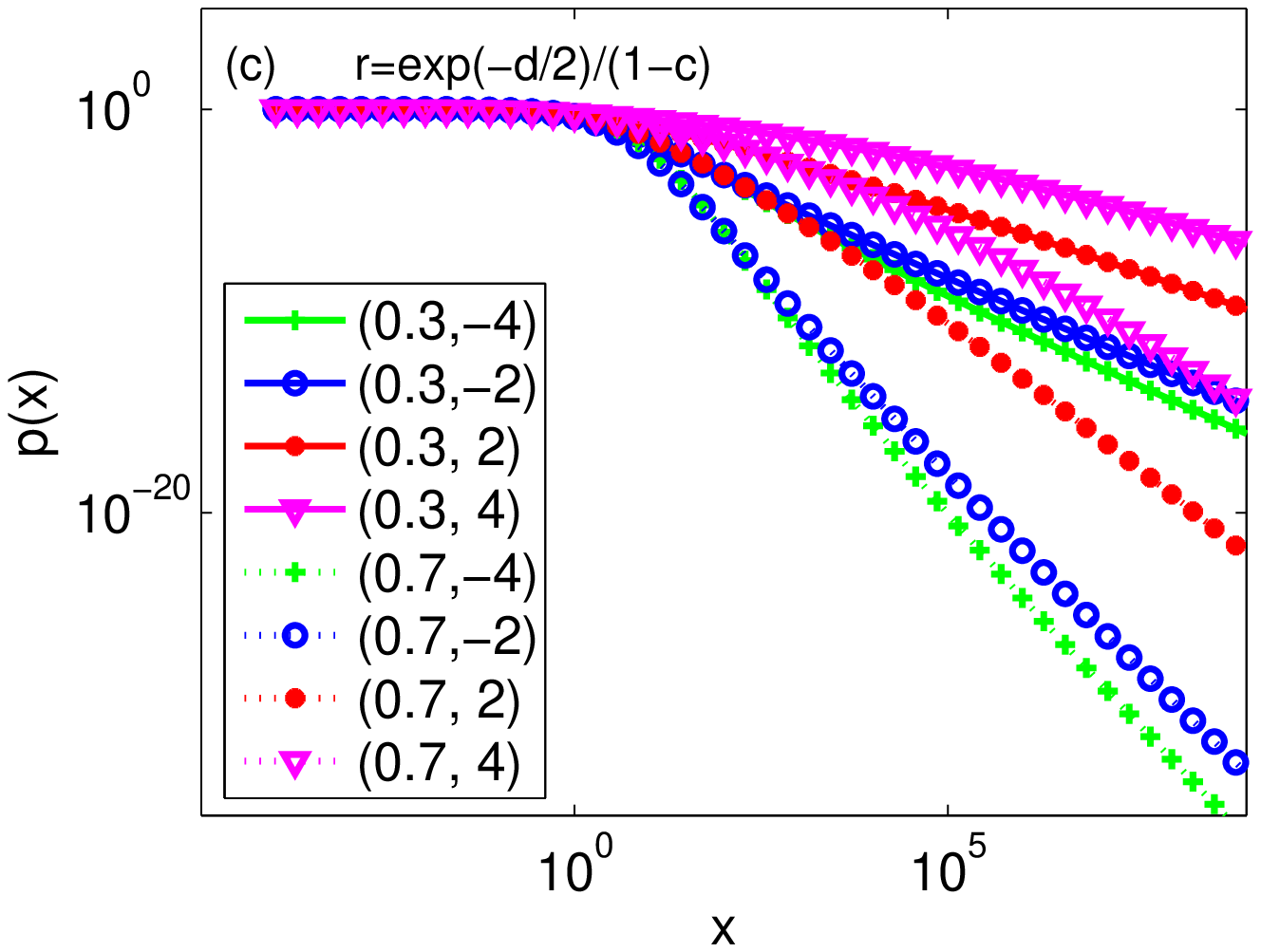}
 \end{center}
\caption{
Distribution functions based on the  'Lambert exponential', $p(x)=\gexp_{c,d,r}(-x)$ are plotted for various $(c,d)$ values.
(a) Asymptotically stable systems, i.e. the stretched exponential limit $c\to 1$. 
It includes the Boltzmann distribution for $d=1$.
(b) The $d\to 0$ limit -- i.e. the $q$-exponential limit. 
In (a) and (b) the black dashed lines represent the stretched exponential ($c=1$) or $q$-exponential ($d=0$)
limit functions.
(c) The general case for distribution functions for various values of $(c,d)$ away from the limits $c\sim 1$ or $d\sim 0$. They should not 
be confused with power-laws. 
\label{MECfig}
}
\end{figure}

\begin{table*}[t]
\caption{
Comparison of several entropies for which $S=\sum_i g(p_i)$, and K1-K3 hold. They are shown as special cases of the entropy given in Eq. (\ref{gent}). Their asymptotic behavior 
is uniquely determined by $c$ and $d$. 
It can be seen immediately that $S_{q>1}$, $S_{b}$ and $S_{E}$ are asymptotically identical. So are $S_{q<1}$ and $S_{\kappa}$ as well as $S_{\eta}$ and $S_{\gamma}$.  
}
\centering
\begin{tabular}{ll|c|c|l}
entropy &  & $c$ & $d$ & reference \\ \hline
$S_{c,d} = er \sum_i \Gamma(d+1, 1-c\ln p_i) - cr $ 	& $(r=(1-c+cd)^{-1})$								& $c$ 		& $d$  			&  \\  \hline
$S_{BG} =  \sum_i p_i  \ln  (1/p_i)    $ 			&												& $1$ 		& $1$  			&   \cite{kinchin_1} \\  \hline
$S_{q<1}(p) = \frac{1-\sum{p_i^q}}{q-1}$ 			& $(q<1)$ 										& $c=q<1$ 	& $0$  			&  \cite{tsallis88} \\  \hline
$S_{\kappa}(p) = - \sum_i p_i \frac{p_i^{\kappa}  - p_i^{-\kappa}  }{2\kappa}  $ 	 &	($0<\kappa \leq1$)						& $c=1-\kappa$ & $0$  			&   \cite{kaniadakis} \\  \hline
$S_{q>1}(p) = \frac{1-\sum{p_i^q}}{q-1}$ 			& $(q>1)$ 										& $1$ 		& $0$  			&   \cite{tsallis88} \\  \hline
$S_{b}(p) = \sum_i (1-e^{-bp_i})   + e^-b -1$ 		& $(b>0)$ 										& $1$ 		& $0$  			&   \cite{curado} \\  \hline
$S_{E}(p) =  \sum_i p_i (1-e^{\frac{p_i -1}{p_i}})    $  &												& $1$ 		& $0 $  			&   \cite{expo_ent} \\  \hline
$S_{\eta}(p) = \sum_i  \Gamma(\frac{\eta+1}{\eta},-\ln p_i) - p_i \Gamma (\frac{\eta+1}{\eta})$  &$ (\eta>0)$ 		& $1$ 		& $d=\frac{1}{\eta}$  	&   \cite{celia} \\  \hline
$S_{\gamma}(p) =  \sum_i p_i \ln ^{1  / \gamma} (1/p_i)    $ 	&										& $1$ 		& $d =1/\gamma $  	&   \cite{TsallisBook_2009}, footnote 11, page 60 \\  \hline
$S_{\beta}(p) =  \sum_i p_i^{\beta} \ln  (1/p_i)    $ 	&												& $c=\beta$  	& $1$  	&    \cite{shafee} \\  \hline

\end{tabular}
\end{table*}

\section{The derivation of entropy}

Since we are dealing with equivalence classes $(c,d)$ we can now look for a 
two-parameter family of entropies, i.e. functions $g_{c,d}$,
such that $g_{c,d}$ is a representative of the class $(c,d)$ for each pair 
$c\in(0,1]$ and $d\in\mathbb{R}$.
A particularly simple choice which covers {\em all} pairs $(c,d)$ is 
\begin{equation}
g_{c,d,r}(x) = r A^{-d}e^{A} \, \Gamma \left(1+d\,,\,A-c\ln x \right)-rcx  \quad, 
\label{gent}
\end{equation}
with $A=\frac{cdr}{1-(1-c)r}$. 
$\Gamma(a,b)=\int_b^\infty dt\,t^{a-1}\exp(-t)$ is the incomplete Gamma-function and 
$r$ is an arbitrary constant $r>0$ (see below). 
For all 
choices of $r$ the function 
$g_{c,d,r}$ is a representative of the class $(c,d)$. 
This allows to choose $r$ as a suitable function of $c$ and $d$. 
For example choose $r=(1-c +cd)^{-1}$, so that $A=1$, and 
\begin{equation}
S_{c,d}[p] = \frac{e \sum_i \Gamma \left(1+d\,,\,1-c\ln p_i \right)}{1-c+cd} - \frac{c}{1-c+cd}  \quad. 
\label{gent2}
\end{equation}
The proof of the correct asymptotic properties is found in SI (Theorem 4).

\section{Special cases of entropic equivalence classes}

Let us look at some specific equivalence classes. 

\begin{itemize}

\item Boltzmann-Gibbs entropy belongs to the $(c,d)=(1,1)$ class. One immediately verifies from Eq. (\ref{gent}) that  
 	\begin{equation}
  	S_{1,1}[p]= \sum_i g_{1,1}(p_i)=   -\sum_i p_i\ln p_i +1  
  	\quad .
 	\end{equation}

\item Tsallis entropy belongs to the  $(c,d)=(c,0)$ class. With Eqs. (\ref{gent}) and (\ref{rrange})
	we get
 	\begin{equation}
		\begin{array}{lcl}
	 		S_{c,0}[p] = \sum_i g_{c,0}(p_i)=  \frac{1-\sum_i p_i^c}{c-1} +1 \, . 
		\end{array}
 	\end{equation}
	Note, that although the {\em pointwise} limit $c\to 1$ of Tsallis entropy is the BG-entropy, the asymptotic properties $(c,0)$ do {\em not} 
	change continuously to $(1,1)$ in this limit! In other words, the thermodynamic limit and the limit $c\to 1$ do not commute.

\item An entropy for stretched exponentials has been given in \cite{celia} which belongs  to the $(c,d)=(1,d)$ classes, see Table 1.
	It is impossible to compute the general case without explicitly using the Gamma-function.   As one specific example  
	we compute the $(c,d)=(1,2)$ case,  
	\begin{equation}
		\begin{array}{lcl}
  		S_{1,2}[p]= 2  \left(1-\sum_i p_i \ln p_i  \right) + \frac{1}{2}\sum_i p_i\left(\ln p_i  \right)^2\\
  	\end{array}
 	\end{equation}
 	The asymptotic behavior is dominated by the second term.  	
	
\item All entropies associated with distributions with  compact support belong to $(c,d)=(1,0)$. 	
Clearly, distribution functions with compact support all have the same trivial asymptotic behavior. 
\end{itemize}
A number  of other entropies which are special cases of our scheme are listed in Table 1.

\section{The distribution functions}

Distribution functions associated with  the $\Gamma$-entropy, Eq. (\ref{gent2}), can be derived from the so-called generalized exponentials, 
$p(\epsilon) = \gexp_{c,d,r}(-\epsilon)$. 
Following \cite{ggent,HT_hilhorst} (see also SI generalized logs), the generalized logarithm $\glog$ 
can be found in closed form
\begin{equation}
\glog_{c,d,r}(x) =      r \, x^{c-1} \,  \left[ 1- \frac{1-(1-c)r}{rd} \ln x  \right]^{d}  \,,
\label{glog}
\end{equation}
and its inverse function, $\gexp=\glog^{-1}$, is 

\begin{equation}
\gexp_{c,d,r}(x)=  e^{    - \frac{d}{1-c}  \left[ W_k \left( B (1-x/r )^{ \frac{1}{d} } \right)  -  W_k(B)  \right]  }\, ,
\label{gexp}
\end{equation}
with the constant $B\equiv \frac{(1-c)r}{1-(1-c)r} \exp \left(  \frac{(1-c)r}{1-(1-c)r} \right) $. 
The function $W_k$ is the $k$'th branch of the Lambert-$W$ function, which is a solution of the equation
$x=W(x)\exp(W(x))$. Only branch $k=0$ and branch $k=-1$ have real solutions $W_k$.
Branch $k=0$ is necessary  for all classes with $d\geq 0$, branch $k=-1$ for $d<0$.

\subsection{Special cases of distribution functions}

It is easy to verify that the class $(c,d)=(1,1)$ leads to Boltzmann distributions, and the class $(c,d)=(c,0)$ yields 
power-laws, or more precisely, Tsallis distributions i.e. $q$-exponentials. 

All classes associated with  $(c,d)=(1,d)$, for $d>0$ are associated with stretched exponential distributions.   
To see it, remember that $d>0$ requires the branch $k=0$ of the Lambert-$W$ function. Using  the expansion $W_0(x)\sim x-x^2+\dots$ for $1\gg|x|$, 
the limit $c\to 1$ turns out  to be a stretched exponential 
\begin{equation}
	\lim_{c\to 1}\gexp_{c,d,r}(x)=e^{-dr\left [\left(1-\frac{x}{r}\right)^{\frac{1}{d}}-1\right]}\quad. 
\label{strexp}
\end{equation}
Clearly,  $r$ does not effect its asymptotic properties, but can be used to modify 
finite size properties of the distribution function on the left side.  
Examples of distribution functions are shown in Fig. \ref{MECfig}. 

\subsection{A note on the parameter $r$}

In Eq. (\ref{gent2}) we chose $r=(1-c+cd)^{-1}$. This is not the most general case. 
More generally,  only the following limitations on $r$ are required if the corresponding generalized logarithms 
(for definition see SI) are wanted to be endowed with the usual 
properties ($\glog(1)=0$ and $\glog'(1)=1$), 
\begin{equation}
\begin{array}{ll}
d>0:& r<\frac{1}{1-c}\quad,\\
d=0:& r=\frac{1}{1-c}\quad,\\
d<0:& r>\frac{1}{1-c}\quad.
\end{array}
\label{rrange}
\end{equation}
Note that every choice of $r$ gives a representative of the equivalence class $(c,d)$,
i.e. $r$ has no effect on the asymptotic (thermodynamic) limit, but it encodes finite-size characteristics.
A particular practical choice for $r$ is $r=(1-c+cd)^{-1}$ for $d>0$ and $r=\exp(-d)/(1-c)$ for $d<0$.

\section{A note on R\'enyi entropy}

R\'enyi entropy is obtained by relaxing K4 to a pure additivity condition, and by  relaxing $S=\sum g$.
For R\'enyi-type entropies, i.e. $S=G(\sum_{i=1}^W g(p_i))$, one gets 
$\lim_{W\to\infty}\hat S(\lambda W)/\hat S(W)= \lim_{s\to\infty} G(\lambda f_g(\lambda^{-1})s)/G(s)$, where 
$f_g(z)=\lim_{x\to 0} g(zx)/g(x)$. 
The expression $f_G(s)\equiv \lim_{s} G(s y)/G(y)$, 
now provides the starting point of a deeper analysis, which follows the same lines as those presented here. 
However, this analysis gets more involved and properties of the entropies get more complicated. 
In particular, R\'enyi entropy, $G(x) \equiv \ln(x)/(1-\alpha)$ and $g(x)\equiv x^{\alpha}$, is
additive, i.e. asymptotic properties, analogous to the ones presented in this paper,
would yield the class $(c,d)=(1,1)$, which is the same as for BG-entropy. However,
R\'enyi entropy can also be shown {\em not} to be Lesche stable \cite{lesche,Abe_2002, Arimitsu,Kaniadakis_2004,HT_robustness_1}. 
This must not be confused with the situation presented above
where entropies were of the form $S=\sum g$. 
All of the $S=\sum g$ entropies can be shown to be Lesche stable 
(see proof SI Theorem 3).

\section{Discussion}

We argued that the physical properties of macroscopic statistical systems being described by 
generalized entropic forms  (of Eq. (\ref{S_g})) can be uniquely classified  in terms of their asymptotic properties  in the limit $W\to\infty$.
These properties are characterized by two exponents $(c,d)$, in nice analogy to critical exponents. 
These exponents  define equivalence relations on the considered classes of entropic forms. 
We showed that a single entropy -- parametrized by the two exponents --  
covers all {\em admissible} systems (Khinchin axioms 1-3 hold, 4 is violated). 
In other words every statistical system has its pair of unique exponents in the large size limit, its entropy is given by 
$S_{c,d} \sim \sum_i  \Gamma \left(1+d\,,\,1-c\ln p_i \right)$ Eq. (\ref{gent2}).

As special cases Boltzmann-Gibbs systems have $(c,d)=(1,1)$, 
systems characterized by stretched exponentials  
belong to the class $(c,d)=(1,d)$, and 
Tsallis systems to $(c,d)=(q,0)$. 
The distribution functions of {\em all} systems $(c,d)$ are  shown 
to belong to a class of  exponentials  involving Lambert-$W$ functions, given in Eq. (\ref{gexp}). 
There are no other options for tails in distribution functions other than these.  

The equivalence classes characterized by the exponents $c$ and $d$, form {\it basins of asymptotic equivalence}.
In general these basins and their representatives will characterize interacting statistical (non-additive) systems.
There is a remarkable analogy between these basins of asymptotic equivalence and
the {\it basin of attraction} of weakly interacting, uncorrelated systems subject to the law of large numbers, i.e. the central limit theorem.
Although, strictly speaking, there is no limit theorem which selects a specific representative within any
such equivalence class, it is clear that any system within a given equivalence class may exhibit
individual peculiarities as long as it  is small. Yet systems of the same class will start behaving
identically as they become larger. Finally, only the asymptotic properties are relevant. 
Distribution functions converge to those functions uniquely determined by $(c,d)$.

Our framework clearly shows that for non-interacting systems $c$ has to be 1.
Setting $\lambda=W_B$ in Eq. (\ref{Stog}) and Eq. (\ref{f_funct}), immediately implies $S(W_AW_B)/S(W_A)\sim  W_B^{1-c}$.
This means that if 
for such a system it would be true that
$c\neq1$,  
then
adding only a few independent states to a system would explosively change 
its entropy
and extensivity would be strongly violated.
A further interesting feature of admissible systems is that they 
all are 
what has been called {\em Lesche stable}. 
systems  (proof in SI Theorem 3). 
As a practical note Lesche stability corresponds one-to-one to the continuity of the scaling function $f$ (see SI) 
and can therefore be checked by a trivial verification of this property (Eq. (\ref{f_funct})). 

We have developed a comprehensive
classification scheme for the generic class of generalized entropic 
forms of type $S=\sum_i g(p_i)$, and commented on how the philosophy 
extends to entropies of e.g. R\'enyi type, i.e. $S=G(\sum_{i} g(p_i))$. 
Finally, we argue that complex statistical systems can be associated with admissible systems  
of equivalence classes $(c,d)$, with $0<c<1$.



\newpage

\section*{\LARGE Supplementary Information}
This supplement to the paper `A classification of complex statistical systems in terms of their stability and a thermodynamical derivation of their entropy and distribution functions' contains detailed information on the technical  aspects of the work. In particular it contains the proofs omitted from the paper for readability.

\section*{Proposition 1}

The consequence of K2 -- that the maximal unconstrained entropy is found for equi-distribution $p_i=1/W$ -- is equivalent 
to the requirement that $g$ is a concave function on $[0,1]$. This is summarized in the well known proposition

{\bf Proposition:}\\
Let $S_g$ be given by Eq. (1) (main text) and let $g$ be a concave function which is continuously differentiable on the semi-open 
interval $(0,1]$ then $\hat S_g(W)\equiv \max_{\sum_i p_i=1} S_g[p]$, is given by $\hat S_g(W)=Wg(1/W)$. 

\begin{proof}
Let $W$ be the number of states $i=1,\dots,W$. The constraint that $p$ is a probability $\sum_{i=1}^W p_i=1$ can be 
added to $S_g$ with by using a Lagrangian multiplier. I.e. differentiation of $S_g[p]-\alpha(\sum_i p_i-1)$ with respect 
to $p_i$ gives $g'(p_i)=\alpha$, where $\alpha$ is the Lagrangian multiplier. Since $g$ is concave $g'$ is monotonically  
decreasing and therefore $p_i=p_j$ for all $i$ and $j$. Consequently $p_i=1/W$ for all $i$ and $\sum_{i=1}^W g(1/W)=Wg(1/W)$.
\end{proof}

\section*{Theorem 1 and proof}

{\bf Theorem 1:} 
Let $g$ be a continuous, concave function on $[0,1]$ with
$g(0)=0$ and 
let $f(z)=\lim_{x\to 0^+}g(zx)/g(x)$ be continuous, then
$f$ is of the form $f(z)=z^c$ with $c\in(0,1]$. 

\begin{proof}
Note that $f(ab)=\lim_{x\to 0}g(ab x)/g(x)=\\ \lim_{x\to 0}(g(ab x)/g(b x))(g(b x)/g(x))=f(a)f(b)$. 
All pathological solutions are excluded by the requirement that $f$ is continuous.
So $f(ab)=f(a)f(b)$ implies that $f(z)=z^c$ is the only possible solution of this equation. 
Further, since $g(0)=0$, also $\lim_{x\to 0}g(0 x)/g(x)=0$, and it follows that $f(0)=0$.
This necessarily implies that $c>0$. $f(z)=z^c$ also has to be concave since $g(z x)/g(x)$ is
concave in $z$ for arbitrarily small, fixed $x>0$. Therefore $c\leq 1$. 
\end{proof}

Note that if $f$ is not required to be continuous, then there are various ways to
construct (rather pathological) functions $f$ solving $f(ab)=f(a)f(b)$ different from $z^c$, as for instance
$f(z)=1$ for $z$ being a rational number and $f(z)=0$ for $z$ being an irrational number,
which is nowhere continuous. Also $f(z)=\lim_{c\to0^-}z^c$, which is zero for $z=0$ and one otherwise, would be a possible
solution.
The continuity requirement eliminates all these possibilities.

\section*{Theorem 2 and proof}

{\bf Theorem 2:}
Let $g$ be like in Theorem 1 and let $f(z)=z^c$ then
$h_c$ given in Eq. (8) is a constant of the form 
$h_c(a)=(1+a)^d$ for some constant $d$.

\begin{proof}
We can determine $h_c(a)$ again by a similar trick as we have used for $f$.
\begin{equation} 
\begin{array}{ll} 
h_c(a)&=\lim_{x\to 0}\frac{g(x^{a+1})}{x^{ac}g(x)}\\
&=\frac{g\left((x^b)^{\left(\frac{a+1}{b}-1\right)+1}\right)}{(x^b)^{\left(\frac{a+1}{b}-1\right)c}g(x^b)}\frac{g(x^{b})}{x^{(b-1)c}g(x)}\\ \nonumber
&=h_c\left(\frac{a+1}{b}-1\right)h_c\left(b-1\right)\quad,
\end{array}
\end{equation}
for some constant $b$.
By a simple transformation of variables, $a=bb'-1$, one gets 
$h_c(bb'-1)=h_c(b-1)h_c(b'-1)$. 
Setting $H(x)=h_c(x-1)$ one again gets $H(bb')=H(b)H(b')$.
So $H(x)=x^d$ for some constant $d$ and consequently 
$h_c(a)$ is of the form $(1+a)^d$.
\end{proof}

\section*{Theorem 3 on Lesche stability and the theorem relating it to continuous $f$, and its proof}

The Lesche stability criterion is a uniform-equi-continuity property of functionals $S[p]$ on
families of probability functions $\{p^{(W)}\}_{W=1}^{\infty}$ where
$p^{(W)}=\{p_i^{W}\}_{i=1}^W$. The criterion is phrased as follows:\\

Let $p^{(W)}$ and $q^{(W)}$ be probabilities on $W$ states. An entropic form $S$ is Lesche stable 
if for all $\epsilon>0$ and all $W$ there is a $\delta>0$ such that
\begin{equation}
||p^{(W)}-q^{(W)}||_1<\delta \Rightarrow |S[p^{(W)}]-S[q^{(W)}]|<\epsilon \hat S(W)\quad, \nonumber
\end{equation}
where $\hat S(W)$ is again the maximal possible entropy for $W$ states.
We now characterize Lesche stability on the class of the generalized entropic forms in terms of the continuity of $f$ in \\

{\bf Theorem 3:}  
Let $p_i\geq0$ be a probability, i.e. $\sum_{i=1}^W p_i=1$, and $W$ the number of states $i$.
Let $g$ be a concave continuous function on $[0,1]$ which is continuously differentiable on the semi-open interval $(0,1]$.
Also, let $g(0)=0$ 
then the entropic form $S_g[p]=\sum_{i=1}^{W} g(p_i)$ is 
Lesche stable iff the function $f(z)=\lim_{x\to 0}g(z x)/g(x)$ is continuous on $z\in[0,1]$. 

\begin{proof} 
Proposition 1 states that the maximal entropy is given by  $\hat S_g(W)=Wg(1/W)$.
We now identify the worst case scenario for $|S_g[p]-S_g[q]|$, where $p$ and $q$ are probabilities on the 
$W$ states. This can be done by maximizing $G[p,q]=|S_g[p]-S_g[q]|-\alpha(\sum_i p_i-1)-\beta(\sum_i q_i-1)-\gamma(\sum_i|p_i-q_i|-\delta)$, 
where $\alpha$, $\beta$ and $\gamma$ are Lagrange multipliers.
Without loss of generality assume that $S_g[p]>S_g[q]$ and therefore the condition $\partial G/\partial p_i=0$ gives
$g'(p_i)+\gamma \, {\rm sign}(p_i-q_i)-\alpha=0$, where $g'$ denotes the first derivative of $g$; ${\rm sign}$ is the signum function.
Similarly, $\partial G/\partial q_i=0$ leads to $g'(q_i)+\gamma \, {\rm sign}(p_i-q_i)+\beta=0$. From this we see that both 
$p$ and $q$ can only possess two values $p_+$, $p_-$ and $q_+$ and $q_-$, where we can assume (without loss of generality)
that $p_+>q_-$ and $q_+>p_-$. We can now assume that for $w$ indices $i$ $p_+=p_i>q_i=q_-$ and for $W-w$ indices $j$
$p_-=p_j<q_j=q_+$ where $w$ may range from $1$ to $W-1$. This leads to seven equations 
\begin{equation}
\begin{array}{lll}
wp_++(W-w)p_-=1 &,& g'(p_+)+\gamma-\alpha=0 \\ 
wq_-+(W-w)q_+=1 &,& g'(p_-)-\gamma-\alpha=0 \\ 
w(p_+-q_-) -    &,& g'(p_+)+\gamma+\beta=0 \\  
-(W-w)(p_--q_+)=\delta &,& g'(p_+)-\gamma+\beta=0  
\nonumber 
\end{array}
\end{equation}
which allow to express $p_-$, $q_-$, and $q_+$ in terms of $p_+$ 
\begin{equation}
\begin{array}{rl}
p_-=&(1-w p_+)/(W-w) \\ 
q_-=&p_+-\delta/2w \\ 
q_+=&(1-w p_+)/(W-w)+\delta/2(W-w)  \quad .
\nonumber 
\end{array}
\end{equation}
Further we get the equation
\begin{equation}
g'(p_+)-g'(p_-)+g'(q_+)-g'(q_-)=0\quad.
\label{impo}
\end{equation}
However, since  $g$ is concave $g'$ is monotonically  decreasing
and therefore $g'(p_+)-g'(q_-)>0$ and $g'(q_+)-g'(p_-)>0$. 
Thus Eq. (\ref{impo}) has no solution, meaning  
that there is no extremum with $p_\pm$ and $q_\pm$ in $(0,1)$, and extrema are at the boundaries. 
The possibilities are $p_+=1$ or $p_-=0$, then $q_+=1$ and $q_-=0$.
Only $p_+=1$ or $p_-=0$ are compatible with the assumption that $S[p]>S[q]$ (the other possibilities are associated
with $S[q]>S[p]$); $p_+=1$ is only a special case of $p_-=0$ with $n=1$. Since $g(0)=0$ this immediately
leads to the inequality
\begin{equation}
\begin{array}{c}
\frac{|S_g[p]-S_g[q]|}{S_{\max}}\leq (1-\phi)\frac{g\left(\frac{\delta}{2(1-\phi)W}\right)}{g(1/W)} 
+ \phi\left|\frac{g\left(\frac{1}{\phi W}\right)}{g(1/W)} -\frac{g\left(\frac{1-\delta/2}{\phi W}\right)}{g(1/W)}\right|
\nonumber
\end{array}
\label{ineq1}
\end{equation}
where $\phi=w/W$ is chosen such that the right hand side of the equation is maximal. 
Obviously for any finite $W$ the right hand side can always be made as small as needed by choosing $\delta>0$
small enough. 
Now take the limit $W\to\infty$. If $f$ is continuous and  using Theorem 1
\begin{equation}
\begin{array}{l}
\frac{|S_g[p]-S_g[q]|}{S_{\max}}\leq \\
\leq (1-\phi)\left(\frac{\delta}{2(1-\phi)}\right)^c  
+ \phi\left|\left(\frac{1}{\phi}\right)^c -\left(\frac{1-\delta/2}{\phi}\right)^c\right|\\ 
\leq (1-\phi)^{1-c}\delta^c 
+ \phi^{1-c}\left|1 -\left(1-\delta/2\right)^c\right|\\
\leq \delta^c + \left|1 -\left(1-c\delta/2\right)\right|\\
\leq \delta^c + \delta\quad.\\
\end{array}
\label{ineq2}
\end{equation}
It follows that $S_g$ is Lesche-stable, 
since we can make the right hand side of Eq. (\ref{ineq2}) smaller than any given $\epsilon>0$ by choosing
$\delta>0$ small enough. This completes the first direction of the proof. 
If, on the other hand, $S_g$ is not Lesche-stable 
then there exists an $\epsilon>0$, such that $|S_g[p]-S_g[q]|/S_{\max}\geq \epsilon$, $\forall N$.
This implies 
\begin{equation}
\begin{array}{l}
(1-\phi)f\left(\frac{\delta}{2(1-\phi)}\right)   
+ \phi\left|f\left(\frac{1}{\phi}\right) -f\left(\frac{1-\delta/2}{\phi}\right)\right|\geq \epsilon \quad , 
\nonumber 
\end{array}
\label{ineq3}
\end{equation}
$\forall \delta>0$.
This again means that either $f(z)$ is discontinuous at $z=1/\phi$ or $\lim_{z\to0}f(z)>0$. 
Since $g(0)=0$ implies that $f(0)=0$, 
$f(z)$ has to be discontinuous at $z=0$. 
\end{proof}

Note that if  $g(x)$ is differentiable at $x=0$, then as a simple Lemma of Theorem 3 
it follows that $S_g$ is 
Lesche stable, since $f(z)=\lim_{x\to 0}g(z x)/g(x)= (g'(0)z x)/(g'(0)x)\to z$.
Consequently, all $g$ analytic on $[0,1]$ are Lesche stable. 
Moreover, $h_1(a)=(g'(0)x^{a+1})/(x^ag'(0)x)=1=(1+a)^0$.
All these $g$ fall into the equivalence class $(c,d)=(1,0)$. 

As an example for a practical use of the above lemma, let us consider a function 
$g(x)\propto\ln(1/x)^{-1}$ for $x\sim 0$. 
Clearly for $z>0$ we have $f(z)=\lim_{x\to 0}g(z x)/g(x)=\ln(x)/\ln(z x)\to 1$. 
On the other hand for $z=0$ we find $f(z)=0$. Therefore, $f(z)$ is not continuous at $z=0$ and 
is violating the preconditions of Theorem 3. 
Lesche instability follows as a Lemma and
does no longer require a lengthly proof.  

\section*{Defining generalized logarithms}\label{app_glog_def}

It is easy to verify that two functions $g_{A}$ and $g_{B}$ 
give rise to equivalent entropic forms, i.e. their asymptotic exponents are identical, 
if $\lim_{x\to 0^+}g_A(x)/g_B(x)=\phi$ and 
$\phi$ is a positive finite constant $\infty>\phi>0$. Therefore
transformations of entropic forms $S_g$ of the type $g(x)\to ag(bx)$, with $a>0$ and $b>0$ positive
constants, lead to equivalent entropic forms.
Following \cite{ggent,HT_hilhorst}, the generalized logarithm $\Lambda_g$ associated with the entropic function
$g$ is basically defined by $-g'(x)$. However, to guarantee that scale transformations of the type
$g(x)\to ag(bx)$ do not change the associated generalized logarithm, $\Lambda_g$, one has to 
define $\Lambda_g(x)=-ag'(bx)$, where constants $a$ and $b$ are fixed
by two conditions 
\begin{equation}
{\rm (i)}\quad \Lambda(1)=0\quad {\rm and}\quad{\rm (ii)} \quad \Lambda'(1)=1\,.
\label{logcond}
\end{equation}
There are several reasons to impose these conditions.
\begin{itemize}
\item
The usual logarithm $\Lambda =\log$ has these properties.
\item
The dual logarithm $\Lambda ^*(x)\equiv -\Lambda(1/x)$ also obeys the the conditions. 
So, if $\Lambda(x)$ can be constructed for $x\in[0,1]$, then $\Lambda $ can be continued to $x>1$ by defining
$\Lambda(x)= \Lambda^*(x)$ for $x>1$ and $\Lambda $ is automatically continuous and differentiable at $x=1$.
\item
If systems $A$ and $B$ have entropic forms $S_{g_A}$ and $S_{g_B}$ which are
considered in a maximum entropy principle then the resulting distribution functions 
$p_{A\,i}={\cal E}_A(-\alpha_A-\beta_A\epsilon_i)$ and
$p_{B\,i}={\cal E}_B(-\alpha_B-\beta_B\epsilon_i)$, where ${\cal E}_{A/B}=\Lambda^{-1}_{A/B}$ are the generalized
exponential functions, then the values of $\alpha$ and $\beta$ of system $A$ and $B$ are directly comparable.
\end{itemize}

Note that to fulfill Eq. (\ref{logcond}) it may become necessary to introduce a 
constant $r$, as we have done in the main text. 

\section*{Proof of asymptotic properties of the Gamma-entropy}\label{app_gamma_asymptotics}

The entropy based on $g_{c,d,r}$, Eq. (10) (main text), indeed  has the desired
asymptotic properties.\\ 

{\bf Theorem 4:} Let $g$ be like in Theorem 3,  
i.e. let $f(z)=z^c$ with $0<c\leq 1$, then
\begin{equation}
\lim_{x\to 0^+}\frac{g'(x)}{\frac{1}{x}g(x)}=c\,.
\label{prehosp}
\end{equation}

\begin{proof}
Consider
\begin{equation}
\begin{array}{lcl}
\lim_{x\to 0^+}\frac{\frac{g(x)-g(zx)}{(1-z)x}}{\frac{1}{x}g(x)}&=&\frac{1}{1-z}\left(\frac{g(x)-g(zx)}{g(x)}\right)\\
&=&\frac{z^c-1}{z-1}\,.
\nonumber
\end{array}
\end{equation}
Taking the limit $z\to 1$ on both sides completes the proof.
\end{proof}
Further, two functions $g_{A}$ and $g_{B}$ 
generate equivalent entropic forms
if $\lim_{x\to 0^+}g_A(x)/g_B(x)=\phi$ and $0<\phi<\infty$. This clearly is true since
\begin{equation}
\begin{array}{lcl}
\lim_{x\to 0^+}\frac{g_A(zx)}{g_A(x)}&=&\frac{g_A(zx)}{g_B(zx)}\frac{g_B(x)}{g_A(x)}\frac{g_B(zx)}{g_B(x)}\\
&=&\phi\phi^{-1}\frac{g_B(zx)}{g_B(x)}\\
&=&\lim_{x\to 0^+}\frac{g_B(zx)}{g_B(x)}\,.
\nonumber
\end{array}
\end{equation}
By an analogous argument the same result can be obtained for the second asymptotic property, Eq. (8) (main text). 
A simple lemma is that given $g_B(x)=ag_A(bx)$, for some suitable constants $a$ and $b$,
then $g_B$ and $g_A$ are equivalent.

A second lemma, following from Eq. (\ref{prehosp}) is that 
\begin{equation}
\lim_{x\to 0^+}\frac{g_A(x)}{g_B(x)}=\lim_{x\to 0^+}\frac{g'_A(x)}{g'_B(x)}\,,
\nonumber
\end{equation}
which is just the rule of L'Hospital shown to hold for the considered families of functions $g$.
This is true since, either $\lim_{x\to 0^+} g_A(x)/g_B(x)=\phi$ with $0<\phi<\infty$ and $c_A=c_B$,
i.e. $g_A$ and $g_B$ are equivalent, or $g_A$ and $g_B$ are inequivalent, i.e.
$c_A\neq c_B$ but $\phi=0$ or $\phi \to \infty$.  

So if one can find a function $g_{\rm test}$, having the desired asymptotic exponents $c$ and $d$, it suffices to show that
$0<-\lim_{x\to0^+}\Lambda_{c,d,r}(x)/g'_{\rm test}(x)<\infty$, where $\Lambda_{c,d,r}$ is the generalized logarithm
Eq. (15) associated with the generalized entropy Eq. (10) (main text).
The test function $g_{\rm test}(x)=x^c\log(1/x)^d$ is of class $(c,d)$, as can be verified easily.
Unfortunately $g_{\rm test}$ can not be used to define the generalized 
entropy due to several technicalities. In particular $g_{\rm test}$ lacks concavity around $x\sim 1$ 
for a considerable range of $(c,d)$ values, which then makes it impossible to define
proper generalized logarithms and generalized exponential functions on the entire interval $x\in[0,1]$. 
However, we only need the asymptotic properties of $g_{\rm test}$ and for 
$x\sim 0$ the function $g_{\rm test}$ does not violate concavity or any other required condition.
The first derivative is $g'_{\rm test}(x)=x^{c-1}\log(1/x)^{d-1}(c\log(1/x)-d)$. 
With this we finally get 
\begin{equation}
\lim_{x\to 0^+}\frac{\Lambda_{c,d,r}(x)}{g'_{\rm test}(x)}=\frac{r-D^{-d}\left(\frac{x}{z}\right)^{c-1}\left(\log\left(\frac{z}{x}\right)\right)^{d}}{x^{c-1}\log(1/x)^{d-1}(c\log(1/x)-d)}
=-\frac{z^{1-c}}{cD^d}\,.
\nonumber
\end{equation}
Since $0<\frac{z^{1-c}}{cD^d}<\infty$ this 
proves that the Gamma-entropy $g_{c,d,r}$, Eq. (10) (main text), represents the equivalence classes $(c,d)$.

\end{document}